\documentclass{nature}

\usepackage{amsmath}
\usepackage{graphicx}
\graphicspath{{figs4/}}

\title{Improved reliability and accuracy of CMIP5 global mean surface temperature projections}

\author{Ehud Strobach$^{1,2}$ \& Golan Bel$^{3,4}$}

\begin{document}

\maketitle

\begin{affiliations}
 \item Earth System Science Interdisciplinary Center, College of Computer, Mathematical, and Natural Sciences, University of Maryland, College Park, MD 20740 USA
 \item Global Modeling and Assimilation Office, NASA Goddard Space Flight Center, Greenbelt, MD 20771 USA
 \item Department of Solar Energy and Environmental Physics, Blaustein Institutes for Desert Research, Ben-Gurion University of the Negev, Sede Boqer Campus 84990, ISRAEL
 \item Center for Nonlinear Studies (CNLS), Theoretical Division, Los Alamos National Laboratory, Los Alamos, NM 87545 USA
\end{affiliations}

\begin{abstract}
Climate predictions are only meaningful if the associated uncertainty is reliably estimated\cite{Murphy_1973,Palmer_2006,Leutbecher_2008}.
A standard practice for providing climate projections is to use an ensemble of projections. The ensemble mean serves as the projection while the ensemble spread is used to estimate the associated uncertainty\cite{Knutti_2013,IPCC_2013}.
The main drawbacks of this approach are the fact that there is no guarantee that the ensemble projections adequately sample the possible future
climate conditions and that the quantification of the ensemble spread relies on assumptions that are not always justified\cite{Tebaldi_2007,Knutti_2013}.
The relation between the true uncertainties associated with projections and ensemble spreads is not fully understood\cite{IPCC_2013}.
Here, we suggest using simulations and measurements of past conditions in order to study both the performance of the ensemble
members\cite{Strobach_2015,Strobach_2016} and the relation between the ensemble spread and the uncertainties associated with their predictions\cite{Strobach_2017b}.
Using an ensemble of CMIP5 long-term climate projections that was weighted according to a sequential learning algorithm and whose
spread was linked to the range of past measurements, we found considerably reduced uncertainty ranges for the projected Global Mean Surface Temperature (GMST).
The results suggest that by employing advanced ensemble methods and using past information, it is possible to provide more reliable and accurate climate projections.
\end{abstract}

Climate predictions (periods of years to decades) and projections (periods of decades to centuries) are usually based on global circulation models, which simulate the multi-scale processes that form the climate system\cite{IPCC_2013}. The complexity of these models and the finite computer power available require parameterization of unresolved processes for any practical simulation\cite{Stensrud_2009}. In addition, the exact state of the system cannot be specified, and there is always some degree of uncertainty associated with the initial condition used in the simulations\cite{Deser_2012}. Therefore, the climate predictions/projections have to be accompanied by the associated uncertainty in order to use them both for scientific research and, obviously, for policy making or practical applications\cite{Tebaldi_2007,IPCC_2013}. One of the simplest methods, which is also commonly used in climate research, is to establish an ensemble of simulations by varying some of the uncertain factors and/or model characteristics (initial condition, parameterization, numerical schemes, grid resolution, model parameters, etc.)\cite{Deque_2007,Smith_2013,Hawkins_2016,Woldemeskel_2016,Strobach_2017}. However, an ensemble of climate simulations is not necessarily representative of the real climate variability; in particular, the ensemble spread cannot be interpreted\cite{Tebaldi_2007,Knutti_2013} as the uncertainty associated with its predictions/projections\cite{Deque_2007,Collins_2013}. Nevertheless, various methods relying on the ensemble spread were used to assess the uncertainty\cite{murphy_2004,IPCC_2013,Knutti_2013,Tebaldi_2011,Power_2012}. The quality of a forecast should be measured by two characteristics: the obvious one is its accuracy (often quantified by the magnitude of the errors), and the second one, which is often overlooked, is its reliability. We refer to the reliability as the correct quantification of the probability of the occurrence of different ranges of conditions\cite{Murphy_1973,Palmer_2006,Leutbecher_2008}.

Recently, a new method for the quantification of the uncertainties associated with ensemble predictions was suggested\cite{Strobach_2017b}. The method is based on studying the relation between the spread of the ensemble member predictions (quantified by the ensemble standard deviation (STD)) and the ensemble root mean squared error (RMSE). Obviously, this approach requires simulations of past conditions, which allow the calculation of the RMSE. The most general method of those suggested\cite{Strobach_2017b} is the Asymmetric Range (AR) method, which relies only on the assumption that the relation between the ensemble spread and the error does not change significantly with time (i.e., the relation found during the learning period remains the same during the prediction/projection period). The prediction of an ensemble is its weighted average (or the simple mean of the ensemble if equally weighted). The AR method has the advantage of estimating independently the range of likely conditions above the mean and the range of conditions below the mean (in this sense, it is asymmetric). The AR method was shown to provide much more reliable predictions (relative to the standard method) for the ensemble of CMIP5 decadal predictions of surface temperature and surface zonal wind\cite{Strobach_2017b}. The improvement of the reliability demonstrated the validity of the assumption that the relation between the ensemble spread and its error does not vary considerably for decadal predictions.

Climate projections are not expected to be synchronized with the natural (or internal) variability of the climate system. However, they are expected to be synchronized with the climate system’s responses to changes in its atmospheric composition.
The reasons for the lack of synchronization include, among others, the turbulent nature of the atmosphere and the oceans (although it develops over different time scales in each of these components) and the initialization method of the projections, which is usually some pseudo-steady state at an arbitrary pre-industrial time\cite{Taylor_2012,IPCC_2013}.
Some synchronization between the simulations and observations can be found in the historical part of the CMIP5 projections (Extended Data Figure \ref{fig:plot_ts_trend_lim_rcp_NCEP_EGA_his}). This synchronization can be attributed to the forcing by the observed atmospheric composition (which mostly varies by the greenhouse gas emissions and large volcanic eruptions; e.g., 1992--1993 cooling related to the Mount Pinatubo eruption\cite{Alan_1995,Parker_1996}, the effects of the Agung eruption in 1963 and the El Chichon eruption in 1982). The climate system responded to the volcanic eruptions within several years. These relatively short response times suggest that comparing the model projections with observations may be used to assess the model performances in simulating the climate system responses to external forcing.

In this study, we used an ensemble of CMIP5 projections\cite{Taylor_2012}. The simulated GMST 20-year running averages for the period of $1967$--$2100$ were considered (for each year, the average GMST of the previous 20 years was considered; running averages could be used in this study because our methodology does not assume that sequential values of the variable are independent). The first stage in the analysis involved weighting the ensemble members according to their past performances (during the learning period of $1967$--$2017$; total of $50$ simulated and observed values of the GMST 20-year average) using the EGA sequential learning algorithm\cite{Strobach_2015,Strobach_2016}. The weighting method is expected to improve the weighted ensemble average forecast\cite{Strobach_2015,Strobach_2016} by finding the optimal combination of weights such that the weighted ensemble mean is as close as possible to the observed value during a learning period.
The weighting affects not only the ensemble mean (the projection of the EGA forecaster) but also the ensemble's STD, which is often used to quantify its spread and the uncertainties associated with the projection.
In the second stage of the analysis, the relation between the weighted STD and the projection errors was established using the AR method\cite{Strobach_2017b} in order to estimate the range of likely GMST values at different periods. The AR method calculates a pair of time-independent correction factors for each desired confidence level (confidence level $c$ implies that the extreme higher and lower tails of the probability distribution, each with a probability $(1-c)/2$, are excluded). The two correction factors ($\gamma_u(c)$ and $\gamma_d(c)$), multiplied by the time-dependent ensemble STD ($\sigma_t$), are then used to determine the uncertainty range for each confidence level: one for the range above the ensemble mean and one for the range below the ensemble mean (see the Methods section). By doing this, the AR method can generate an asymmetric interval for selected confidence levels. The performance of our methodology was tested by splitting the learning period into learning and validation periods. A 35-year learning period (15-year validation period) was found to be long enough to reduce the projection error by $25\%$, to decrease the $0.9$ confidence level uncertainty range by $25\%$, and to be more reliable (see also the Methods section and Extended Data Figure \ref{fig:plot_ts_trend_lim_rcp_NCEP_EGA_his}).

\begin{figure}
\begin{center}
\includegraphics[width=39pc]{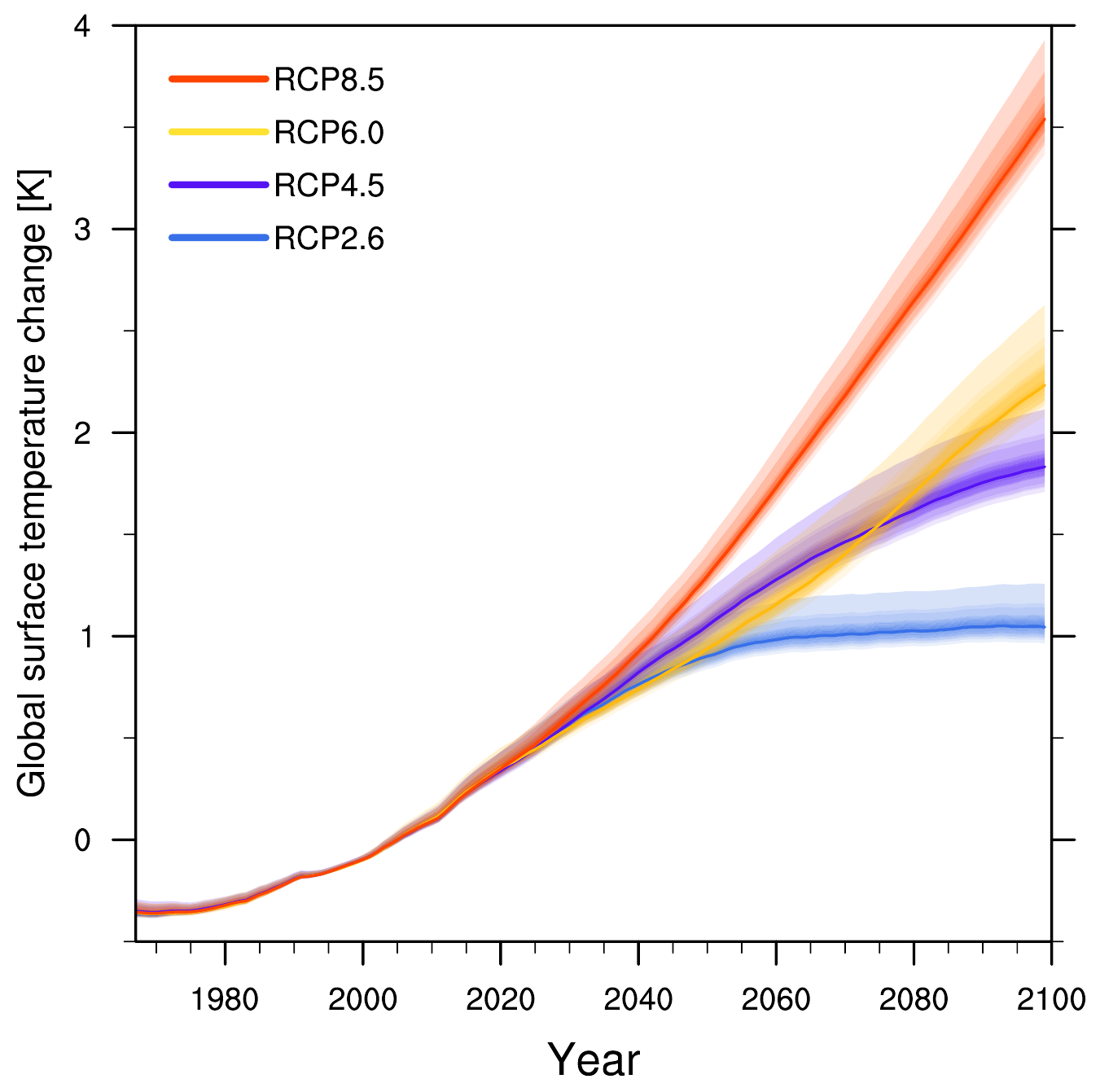}
\caption{\label{fig:plot_ts_trend_lim_rcp_NCEP_EGA} The 20-year running average of Global Mean Surface Temperature (GMST) change relative to the $1986$--$2005$ average for the RCP scenarios included in the CMIP5. The thick lines represent the weighted ensemble mean for the 20-year running average GMST projections, and the shadings represent different significance levels (from $0.1-0.9$) of the associated uncertainty (based on the EGA weighted ensemble and the AR estimation of the uncertainty ranges).}
\end{center}
\end{figure}

Figure \ref{fig:plot_ts_trend_lim_rcp_NCEP_EGA} shows the GMST uncertainties for the different RCP scenarios. Compared to the simple averages (Extended Data Figure \ref{fig:plot_ts_trend_lim_rcp_NCEP_AVG_Gaussian}), the uncertainty ranges in this figure are considerably reduced by using the learning process (learning the relation between the ensemble spread and its error).
The uncertainty range of the $0.9$ confidence level was found to be $68\%$--$78\%$ smaller than the range calculated using an equally weighted ensemble and assuming a Gaussian distribution of the ensemble projections. Similar ratios, between the uncertainty ranges estimated using the EGA and AR methods and those estimated using the equally weighted ensemble and the Gaussian assumption, were also found for other confidence levels (see Extended Data Table \ref{quantile_AR/GS}; numerical values of the ranges estimated using the two methods can be found in Extended Data Tables \ref{quantile_ranges} and \ref{quantile_ranges2}).

We also found that the distribution of the GMST 20-year average is highly asymmetric. For example, the higher significance levels are skewed toward higher than the mean values, and for the $0.9$ confidence level, the $\gamma_u$ values are even more than twice as large as the $\gamma_d$ values (namely, the range above the ensemble mean including $45\%$ of the probability is more than twice as large as the range below the mean, which includes the same probability). The values of $\gamma_u$ and $\gamma_d$ for different significance levels and RCPs are given in Extended Data Table \ref{gammas}. In Extended Data Table \ref{skewkurt}, we provide the skewness and the excess kurtosis of the distribution of the 20-year average GMST. As can be seen, for all RCPs, the skewness does not vanish and is positive (implying that the distribution is right-tailed), and the excess kurtosis is positive, implying a distribution for which rare events are more likely than in a Gaussian distribution.
These results demonstrate that the AR method is not only capable of reducing the uncertainty ranges but is also capable of extracting the deviations from a Gaussian distribution and, therefore, provides more accurate and reliable estimates of the GMST probability distribution.

\begin{figure}
\begin{center}
\includegraphics[width=38pc]{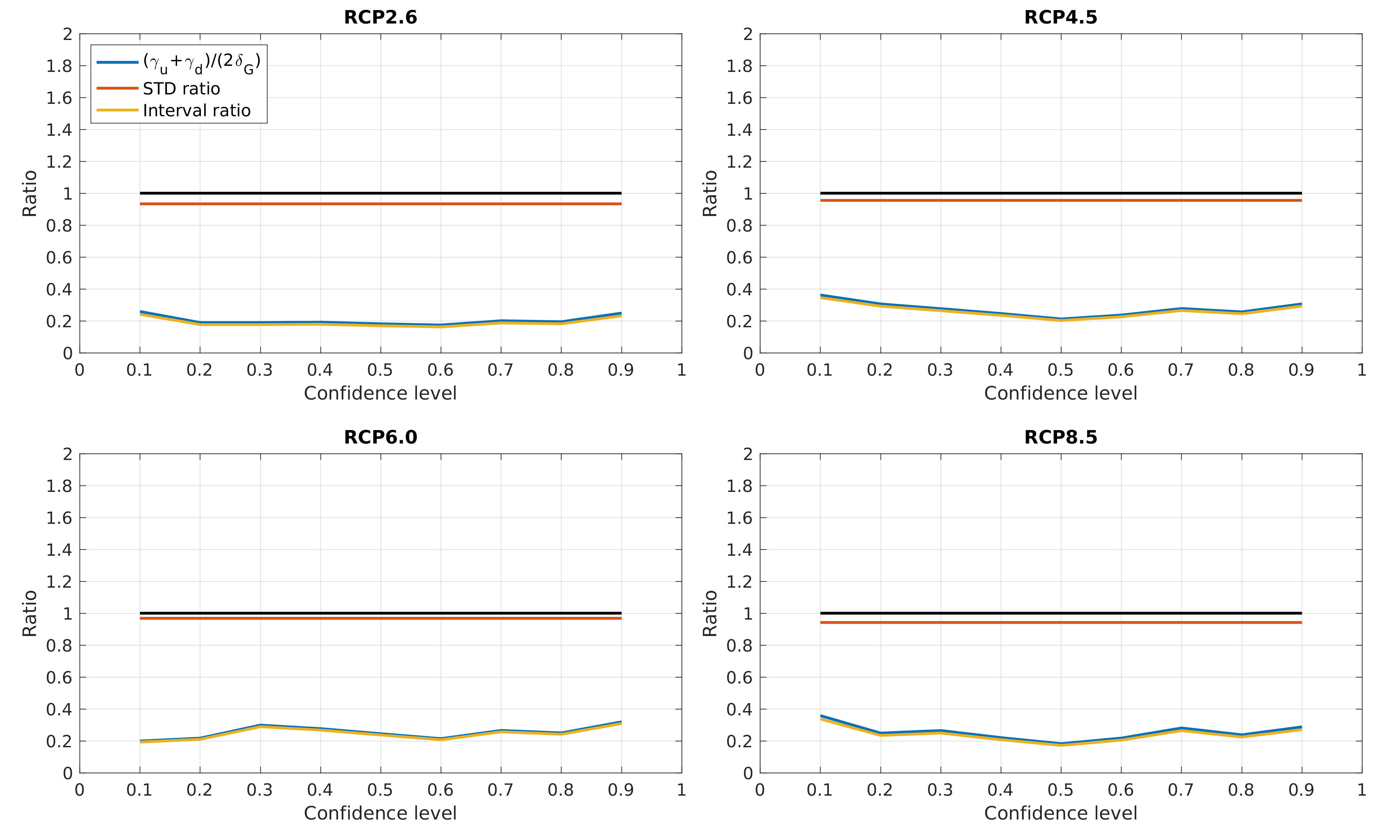}
\caption{\label{fig:cal} The ratio between the confidence intervals of the AR and Gaussian methods for different confidence levels. The difference in the intervals stems from the different STDs between the EGA and equally weighted ensembles and also from the correction factors due to the non-Gaussian distribution of the ensemble projections. The orange line presents the ratio between the EGA and equally weighted ensemble STDs (note that it differs between the RCPs due to the different ensembles and weights); the yellow curve represents the ratio between the sum of the AR correction factors ($\gamma_u(c)+\gamma_d(c)$) and the expected sum of the Gaussian distribution correction factors for a given confidence level ($2 \cdot \delta_G(c)$); and the blue curve represents the total uncertainty reduction, namely, the ratio between the uncertainty ranges estimated using the EGA and AR methods and those estimated using the equally weighted ensemble and the assumption of a Gaussian distribution of the ensemble projections, for different confidence levels.}
\end{center}
\end{figure}

There are two elements that may influence the estimated uncertainty ranges using the EGA and AR methods: the EGA weights, which affect the ensemble STD (weighted STD vs. simple STD), and the AR method correction factors, which multiply the STD and provide the relation between the STD and the uncertainty range for different confidence levels. The first element is time-dependent, and the second is constant during the projection period (see the Methods section for a more detailed discussion). We found that the average ($2020$--$2099$) EGA weighted STD is similar to the average STD of the equally weighted ensemble (the orange lines in Figure \ref{fig:cal}). This suggests that the EGA learning did not converge to one specific model but rather spread the weights among different models. Specifically, we found that only the MRI-CGCM3 has a considerably larger weight than the others, but most of the models were assigned a non-negligible weight (see Extended Data Tables \ref{Models_table} and \ref{Models_weights}).
The main uncertainty reduction is due to the AR method ($\frac{\gamma_{u}(c)+\gamma_{d}(c)}{2\cdot \delta_G(c)}<1$; where, $2\cdot \delta_G(c)$ is the correction factor corresponding to a Gaussian distribution). We also found that this reduction is true not only for the temporal average ($2020$--$2099$) but also for the entire time series of the projections (see Extended Data Figure \ref{fig:cal_std_ratio_sym}).

\begin{figure}
\begin{center}
\includegraphics[width=37pc]{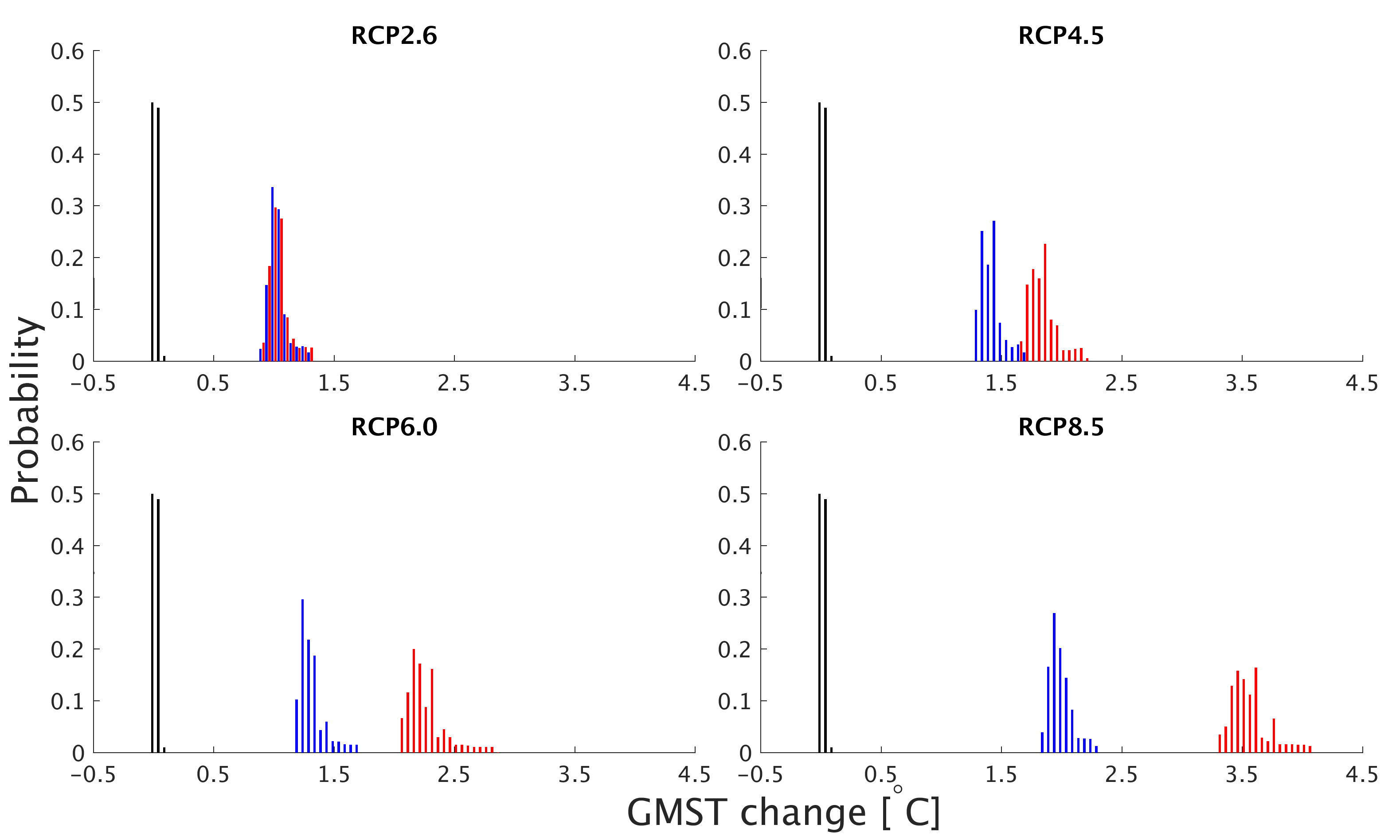} \\
\includegraphics[width=16pc,trim=1.cm 0.6cm 1.5cm 0.5cm, clip]{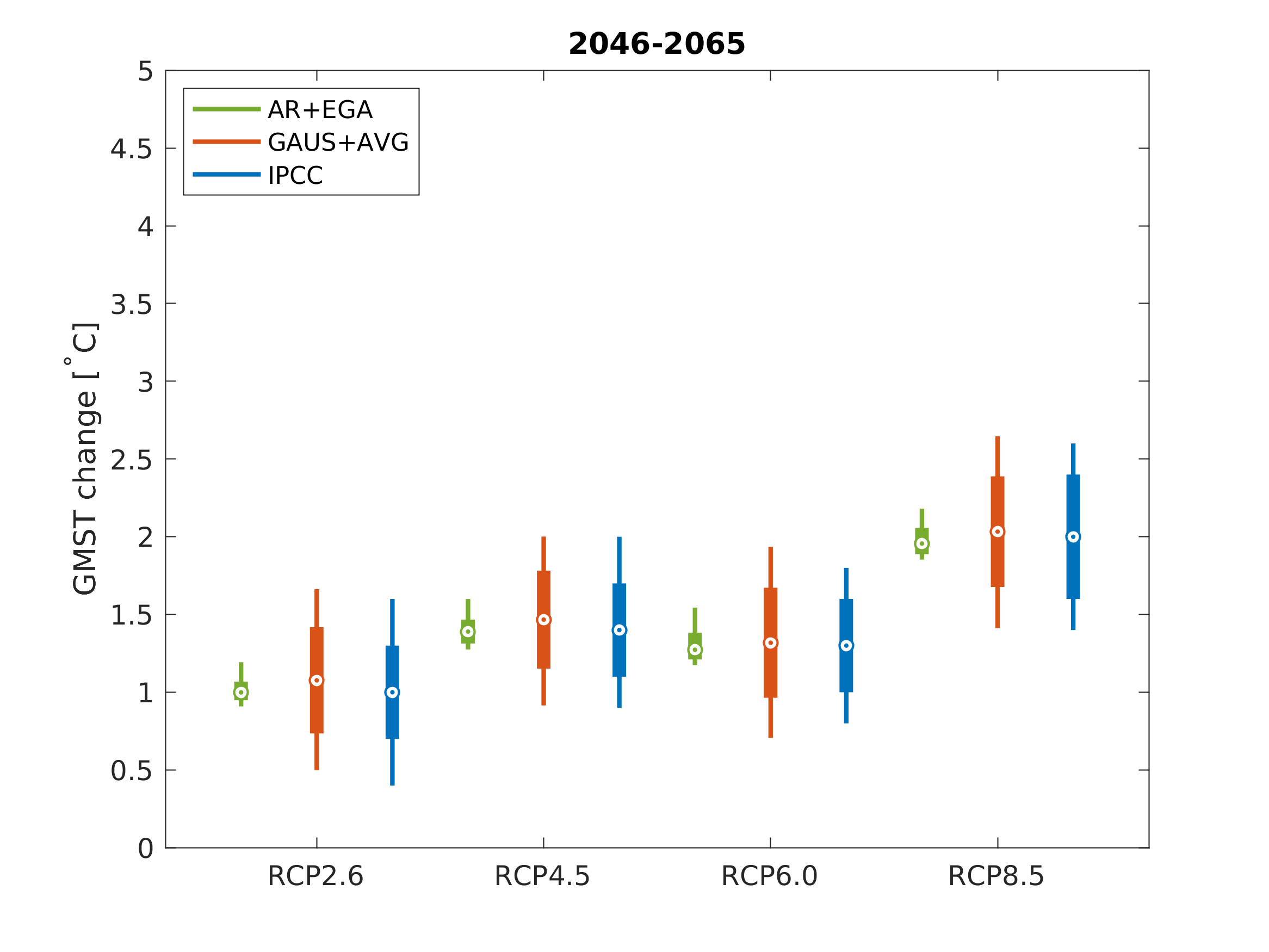}
\includegraphics[width=16pc,trim=1.cm 0.6cm 1.5cm 0.5cm, clip]{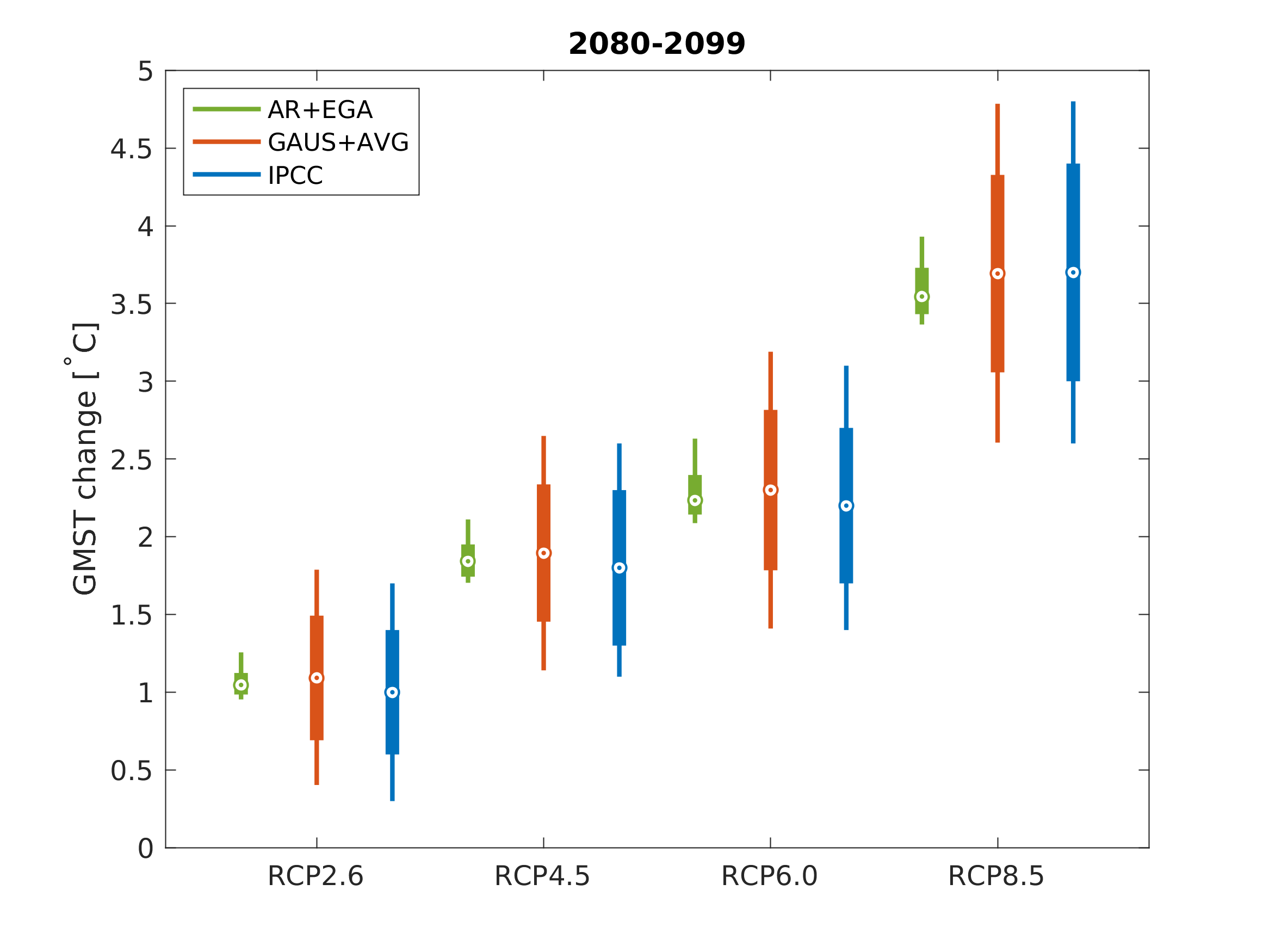}
\caption{\label{fig:pdf_year_combine} \textbf{Four upper panels:} The GMST change probability distributions of the $2046$--$2065$ average (blue) and $2080$--$2099$ average (red) for the four RCPs included in CMIP5, relative to the $1986$--$2005$ NCEP reanalysis average GMST (black). \textbf{Two lower panels:} Box plots of the GMST change relative to the $1986$--$2005$ NCEP reanalysis average. The circles, boxes and error bars represent the ensemble mean, the $16.7\%$--$83.3\%$ uncertainty range, and the $5\%$--$95\%$ uncertainty range, respectively. The blue, red and green colors correspond to the IPCC, the estimation based on the equally weighted ensemble and the Gaussian assumption, and the estimation based on the EGA and AR methods, respectively.}
\end{center}
\end{figure}

The PDs (probability distributions) of the GMST in different years differ due to the temporally varying STD, $\sigma_t$, and mean of the ensemble of GMST projections.
In Figure \ref{fig:pdf_year_combine}, we present the probability distribution of the change in the 20-year average GMST, relative to the NCEP reanalysis $1986$--$2005$ average, for two different periods and for the four RCPs included in the CMIP5.
All RCPs predict significantly warmer GMST in the future, which is indicated by the separation of the $1986$--$2005$ average GMST probability distribution from the probability distributions of the $2046$--$2065$ and $2080$--$2099$ average GMSTs. As expected, the uncertainties also increase with the increased lead time of the projections (indicated by the broadening of the distributions, see also Extended Data Tables \ref{quantiles_20yr} and \ref{quantiles_ranges_20yr}). We also note that the larger the assigned change in the greenhouse gas concentration, the larger is the growth of the uncertainty. Relative to the ranges provided in the last IPCC report\cite{IPCC_2013}, our estimates of the uncertainty ranges are considerably reduced for all the scenarios (Extended Data Tables \ref{quantiles_20yr} and \ref{quantiles_ranges_20yr}).

Uncertainties in climate projections are of great importance for policy makers and practical applications including the development of adaptation and mitigation plans. The adequate quantification of the uncertainties and their reduction, where possible, are also at the core of climate dynamics research.
The fundamental assumption underlying our methods is that it is legitimate to use our knowledge regarding past conditions and the simulations of these conditions in order to learn the relation between them.
Dividing the historical period into learning validation periods revealed that for the conditions in the last century, the assumption is valid.
The combination of the EGA learning algorithm (to weight the models) and the AR method (to extract the relation between the ensemble spread and the actual uncertainties associated with the weighted ensemble projection) resulted in a considerable reduction of the uncertainties for all the RCPs and for the entire period of the projection. The reduction for the 20-year averages reached over $80\%$. Moreover, the entire probability distribution was derived by considering different confidence levels.
A comparison of the CMIP5 ensemble spread with past observations clearly reveals an underconfident ensemble projection, suggesting that the actual uncertainty associated with these projections may be even smaller than estimated here.
We propose that the methods used here should find wider application in the analysis of ensembles of climate predictions.
\begin{methods}

\subsection{Models and data}

We used all available surface temperature projections from the CMIP5 data portal. Projections of some of the models that were included in the last IPCC assessment report were not available when this study was initiated. Several other models were excluded from this study either because of incomplete data for the simulated period spanned in this study or because the cell area information was missing (for the calculation of area-weighted global means). Extended Data Table \ref{Models_table} lists the models that were included in the ensembles considered in this study and their scenario availability (different ensembles were constructed for each RCP based on the data availability).
The internal variability of climate projections was found to be small compared with the model variability\cite{Hawkins_2009}, and therefore, we decided to use one realization (initial condition) per model (the first listed in the CMIP5 data portal).
The performance of each model was assessed by comparing the simulated GMST with the NCEP reanalysis data\cite{Kalnay_1996}, which was considered here as the true value. The NCEP reanalysis was chosen here to represent the true value because it is a widely accepted reanalysis product and because of its relatively long record starting from $1948$; the JRA55 reanalysis\cite{Kobayashi_2015} was also tested but the fact that this reanalysis is shorter, thereby only allowing a shorter learning period, resulted in considerably larger uncertainties. We also tested observation-based products such as the HadCRUT4\cite{Morice_2012} and GISTEMP v3\cite{Hansen_2010} global temperature datasets. These products resulted in similar uncertainty ranges as the NCEP, and therefore, they are not shown here.
The simulations used in this study spanned the period between $1948$, the first year for which NCEP/NCAR reanalysis data are available, and $2100$. Both simulated results and the NCEP/NCAR reanalysis were time averaged to 20-year running averages (the value for each year in the time series corresponds to the average of the year and the 19 preceding years’ GMST, thereby providing a time series from 1967 to 2100).

\subsection{The weighting of the ensemble models}
To weight the climate models, we used the Exponentiated Gradient Average (EGA) algorithm\cite{cesa_2006,Strobach_2015,Strobach_2016}. The inputs to the EGA algorithm are a time series of simulated GMST from an ensemble of ``experts'' (the climate models) and a time series of the ``true value'' (observations; in this study, we used the NCEP/NCAR reanalysis data as the observations). The EGA algorithm compares the simulated results from each of the ensemble members with past observations (using a squared error metric in our study) and weights the models based on their past performances.
As a preliminary step, we bias-corrected the CMIP5 model outputs by subtracting from each model its temporal average during the learning period ($1967$--$2017$) and adding to it the NCEP reanalysis temporal average for the same period. The input to the EGA algorithm is, therefore, bias-corrected CMIP5 projections and NCEP reanalysis ``observations.'' The output of the EGA at the end of the learning period is a weight for each model in the ensemble (Extended Data Table \ref{Models_weights} shows the resulting weights for each model and RCP scenario). The original method was modified to ensure that there are no large fluctuations in the weights during the learning period and that the learning rate is optimal\cite{Strobach_2015,Strobach_2016}.
The weighting procedure allowed us to derive the weighted ensemble STD, which in turn was used to derive the relation between the ensemble spread and the actual uncertainty range based on the Asymmetric Range (AR) method\cite{Strobach_2017b}.
The comparison of simulated results with an observation-based product in this study does not suggest that we are trying to predict the future natural variability of the climate system (such as El Ni{\~n}o events); rather, it suggests that we weight the models based on their ability to correctly simulate the response of the climate system to observed forcing fluctuations over longer time scales. It is worth noting that repeating the same analysis using the annual averages rather than the 20-year averages also resulted in reduced uncertainties. However, in order to avoid criticism of the use of annual averages that are not expected to be synchronized with the simulated dynamics, we focus here on the 20-year averages.

\subsection{Constructing the PD of the GMST} The AR method uses the past errors (relative to the NCEP/NCAR reanalysis) and ensemble STD to construct the PD of the GMST by multiplying the time-dependent (EGA or equally weighted) ensemble standard deviation (STD) with two optimized, significance-level-dependent correction factors that are time-independent. There are two correction factors for each significance level, one for the upper (above the mean) side of the PD ($\gamma_u(c)$) and one for the lower side (below the mean) of the PD ($\gamma_d(c)$), to allow for non-symmetric PDs to be captured. The upper and lower correction coefficients, $\gamma_{u,d}(c)$, are calculated after learning the fraction of the number of observations within a specific range $\big(p_t-\gamma_d(c) \cdot \sigma_t \big) - \big(p_t + \gamma_u(c) \cdot \sigma_t \big)$ during the learning period. The values of $\gamma_{u,d}(c)$ are chosen to be the smallest values that satisfy the conditions that at least a fraction $c/2$ of the observations are inside the area spanned by the two time series $\left[p_t,\gamma_u(c) \cdot \sigma(t)\right]$ during the learning period and similarly the fraction of observations within the area between the two time series $\left[\gamma_d(c) \cdot \sigma(t),p_t\right]$ is $c/2$. Mathematically, these conditions are described by the following equations:
\begin{equation}
\label{eq1}
\begin{aligned}
\gamma_u(c) = \inf \lbrace \gamma_u \in \Re_{>0} : \frac{1}{n}\sum_{t=1}^n \bigg(  \Theta \Big[ \Big(p_t + \gamma_u \cdot \sigma_t \Big)-y_t \Big] \bigg)  \geq \Big( \frac{1+c}{2} \Big) \rbrace
\end{aligned}
\end{equation}
\begin{equation}
\label{eq2}
\begin{aligned}
\gamma_d(c) = \inf \lbrace \gamma_d  \in \Re_{>0} : \frac{1}{n}\sum_{t=1}^n \bigg( \Theta \Big[ y_t-\Big(p_t - \gamma_d \cdot \sigma_t \Big) \Big] \bigg) \geq \Big( \frac{1+c}{2} \Big) \rbrace
\end{aligned}
\end{equation}
In the above equations, $\Theta(x)$ is the Heaviside step function ($\Theta(x)=1$ for $x>0$, $\Theta(x)=0$ for $x<0$, and $\Theta(0)=1/2$), $p_t$ is the weighted ensemble average (forecasts for time $t$), $\sigma_t$ is the weighted ensemble STD at time $t$, $y_t$ is the ``observed'' (``true``) value at time $t$, and $n$ is the number of time points (length of the time series used) in the learning period. For example, $\gamma_{u,d}(c)$ should both be equal to the \textit{probit} function ($\delta_G =\sqrt{2} \mathit{erf}^{-1}(1-c)$), if the observed distribution of the error is unbiased, and Gaussian. We repeated this process for multiple significance levels between $0.1$ and $0.9$.
The resolution of the derived PD depends on the number of observations with values higher and lower than the ensemble projection during the learning period. The optimization of the correction coefficients $\gamma_{u,d}(c)$ can be done by including or excluding at least one observation, and this in turn limits the PD resolution to $1/N_{u,d}$ (where $N_{u,d}$ are the numbers of observations larger and smaller than the projection, respectively; in this study, it was $N_u=24$ and $N_d=26$) in the above and below mean sides, respectively. Note that according to eqs. (\ref{eq1}) and (\ref{eq2}), the actual confidence level might be higher than the desired one due to the finite resolution.
The AR method was developed initially for decadal climate predictions in which it was assumed (and verified) that the uncertainty correction coefficients ($\gamma_{u,d}(c)$) show only small fluctuations during the learning and the validation periods.

\subsection{20-year average probability distributions} The above analysis results in the time series of the mean and the STD (based on the EGA weighted or equally weighted ensembles) and the values of $\gamma_{u,d}(c)$ for a range of desired confidence levels. As outlined in eqs. (\ref{eq1}) and (\ref{eq2}), for each confidence level, the excluded tails on both sides of the mean are equal (the integral over each of the tails is $(1-c)/2$). Considering the quantities above allows one to construct the full probability distribution. The AR algorithm outputs are ranges as a function of probabilities. To convert to probabilities as a function of ranges, we drew $10^7$ values from the derived distribution of the 20-year average GMST of $2065$ and $2099$. The depicted PDs of the 20-year average GMST are the histograms of the $10^7$ values.
We verified that the probability distribution converges for this number of realizations (the differences were below any statistical significance).

\subsection{Testing the EGA and AR performance} We tested the performance of our methodology by dividing the $50$ year period of $1967$--$2017$ into two periods: a learning period and a validation (prediction) period. We tested different combinations of learning and validation periods, and we found that in order to improve the forecast (in terms of accuracy and reliability), we needed more than 30 years of learning. In Extended Data Figure \ref{fig:plot_ts_trend_lim_rcp_NCEP_EGA_his}, we show the forecast from $35$ years of learning ($1967$--$2002$) and $15$ years of a validation experiment 
(in all CMIP5 projections, the assigned atmospheric composition for the historic part until $2005$ is based on observation, and later on, it depends on the RCP. We used the ensemble for RCP 4.5 and used the validation period of $2003$--$2017$, which includes years with RCP rather than measurements based on the atmospheric composition in the last $12$ years ($2006$--$2017$); it is worth mentioning that the variability between the different scenarios is small compared with the model and internal variabilities during the first years of the projections\cite{Hawkins_2009}).
We found that the RMSE of the equally weighted ensemble is larger than that of the EGA weighted ensemble ($0.050^\circ$C for the simple average compared to $0.038^\circ$C for the EGA weighted average); in addition, using the AR method for estimating the uncertainty ranges resulted in smaller future uncertainty ranges for the EGA weighted ensemble and more reliable predictions (Extended Data Figure \ref{fig:plot_ts_trend_lim_rcp_NCEP_EGA_his}).
For the equally weighted ensemble and the assumption of a Gaussian distribution of the ensemble projections, we found that the uncertainty range was much larger than the range expected from a reliable ensemble (the projections using these methods represented an underconfident forecast). A forecast based on the EGA weighted ensemble combined with the AR method was found to be close to reliable (see also Extended Data Fig. \ref{fig:plot_ts_trend_lim_rcp_NCEP_EGA_his}).
Due to the better performance of the EGA weighted ensemble combined with the AR method, we focused on the results of this methodology.

\end{methods}

\bibliography{ref.bib}

%\bibitem{dummy} Articles are restricted to 50 references, Letters to 30.

%% Here is the endmatter stuff: Supplementary Info, etc.
%% Use \item's to separate, default label is "Acknowledgements"

\begin{addendum}
 \item We acknowledge the World Climate Research Programme's Working Group on Coupled Modelling, which is responsible for CMIP, and we thank the climate modeling groups (listed in Extended Data Table \ref{Models_table} of this paper) for producing and making available their model output. For CMIP the U.S. Department of Energy's Program for Climate Model Diagnosis and Intercomparison provides coordinating support and led development of software infrastructure in partnership with the Global Organization for Earth System Science Portals.
 \item[Competing Interests] The authors declare that they have no competing financial interests.
 \item[Correspondence] Correspondence and requests for materials should be addressed to \\ E.S.~(email: strobach@umd.edu).
 \item[Author contributions] Both authors contributed to the design of the research and the writing of the manuscript. ES performed the analysis and generated the figures.

\setcounter{figure}{0}
\renewcommand{\figurename}{Extended Data Figure}
\begin{figure}
\begin{center}
\includegraphics[width=38pc,trim={0 0 0 0},clip]{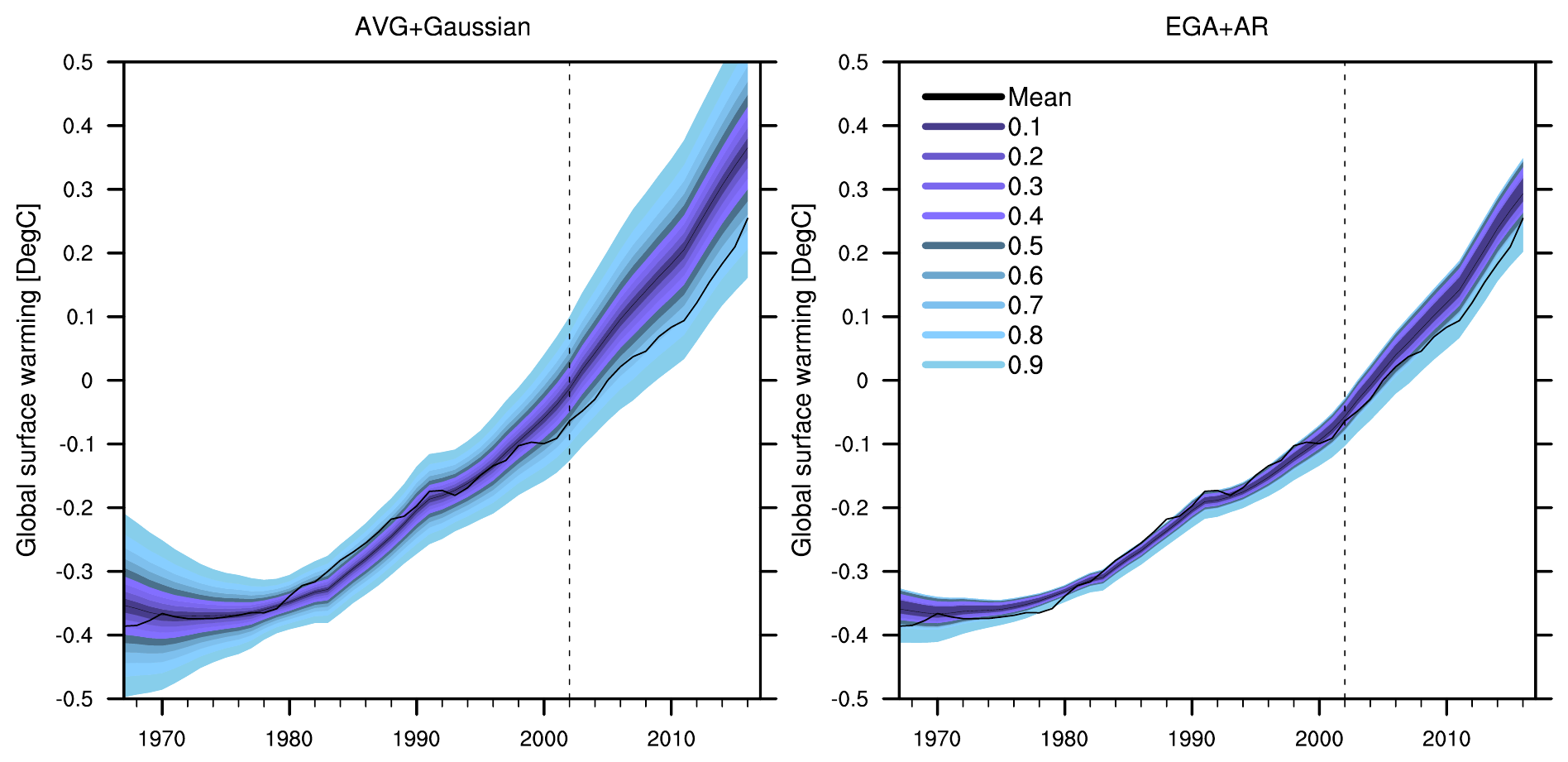}
\includegraphics[width=19pc,trim={0 0 0 0},clip]{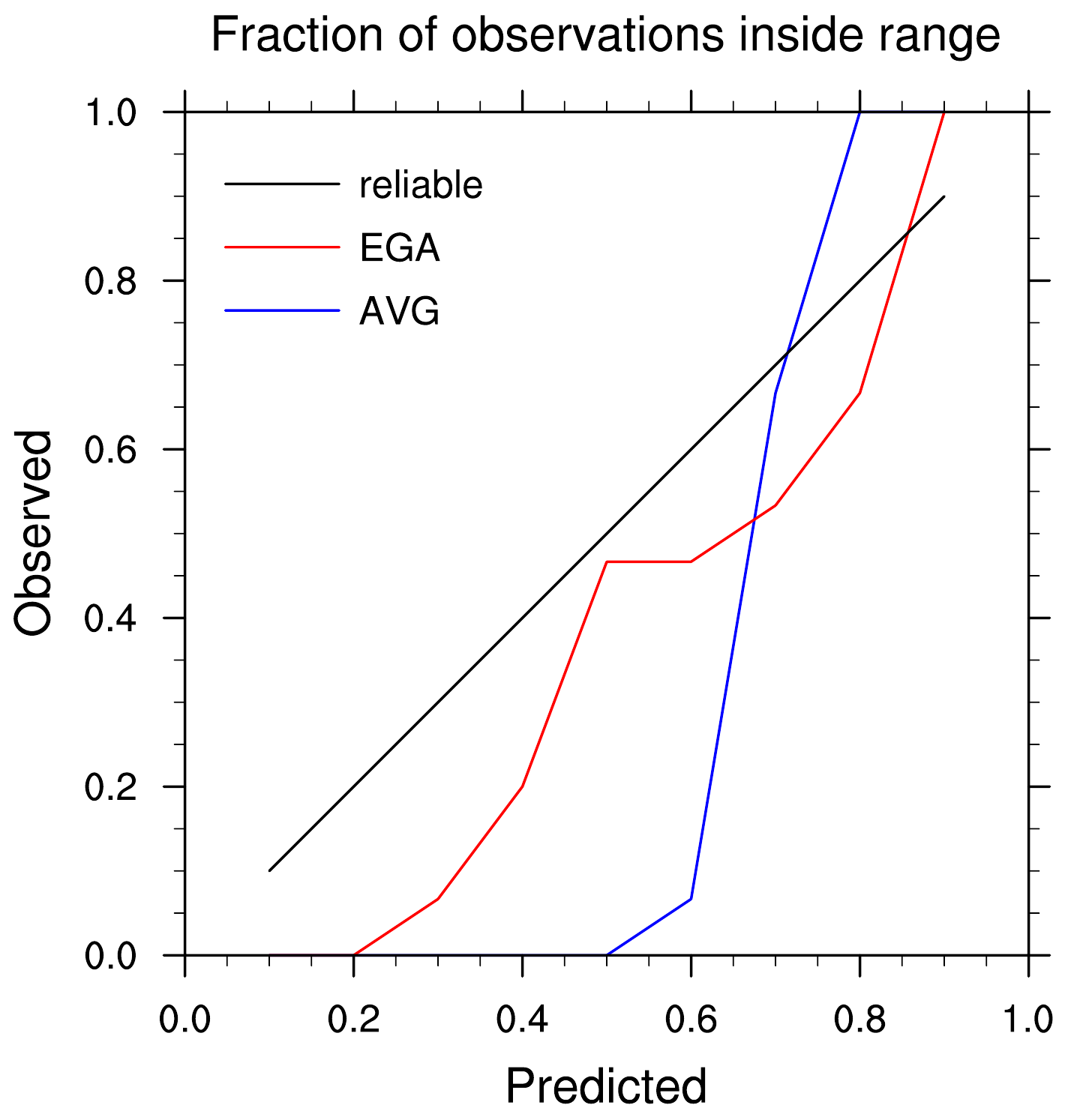}
\caption{\label{fig:plot_ts_trend_lim_rcp_NCEP_EGA_his} \textbf{Top panels:} Global mean surface temperature change relative to $1986$--$2005$ average. The black lines represent the ensemble mean, and the shadings represent different significance levels ($0.1$--$0.9$). The left part of each panel (to the left of the dashed vertical line) represents the learning period, and the right part of each panel represents the validation period. \textbf{Bottom panel:} Prediction period reliability diagram (the fraction of points within the estimated range vs. the expected fraction).}
\end{center}
\end{figure}

\begin{figure}
\begin{center}
\includegraphics[width=38pc]{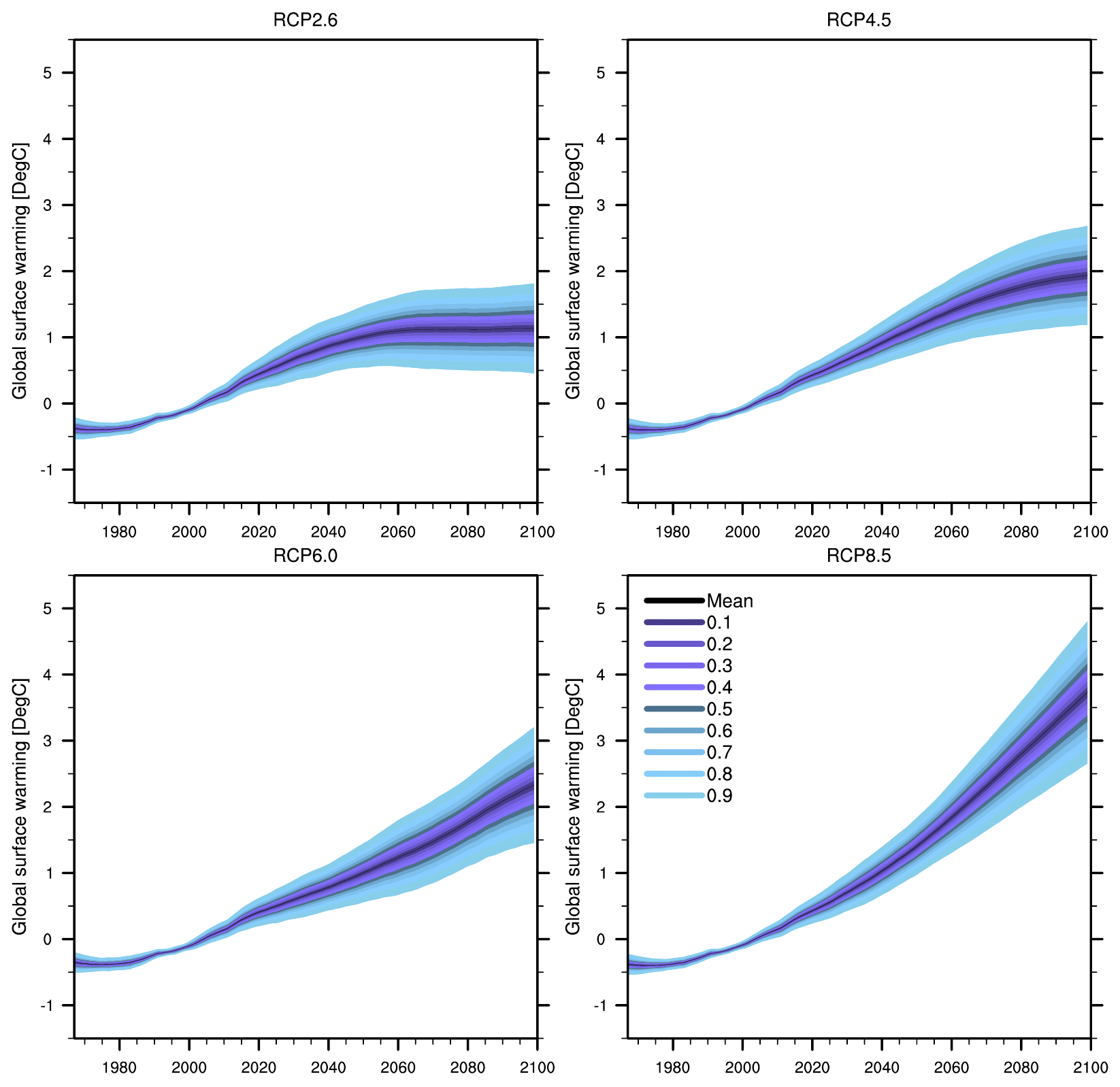}
\caption{\label{fig:plot_ts_trend_lim_rcp_NCEP_AVG_Gaussian} Global mean surface temperature (GMST) change relative to the $1986$--$2005$ average for the RCP scenarios included in CMIP5. The black lines represent the ensemble mean for the 20-year average GMST, and the shadings represent the uncertainty range, for different significance levels, based on the Gaussian assumption and equally weighted ensemble.}
\end{center}
\end{figure}

\begin{figure}
\begin{center}
\includegraphics[width=38pc]{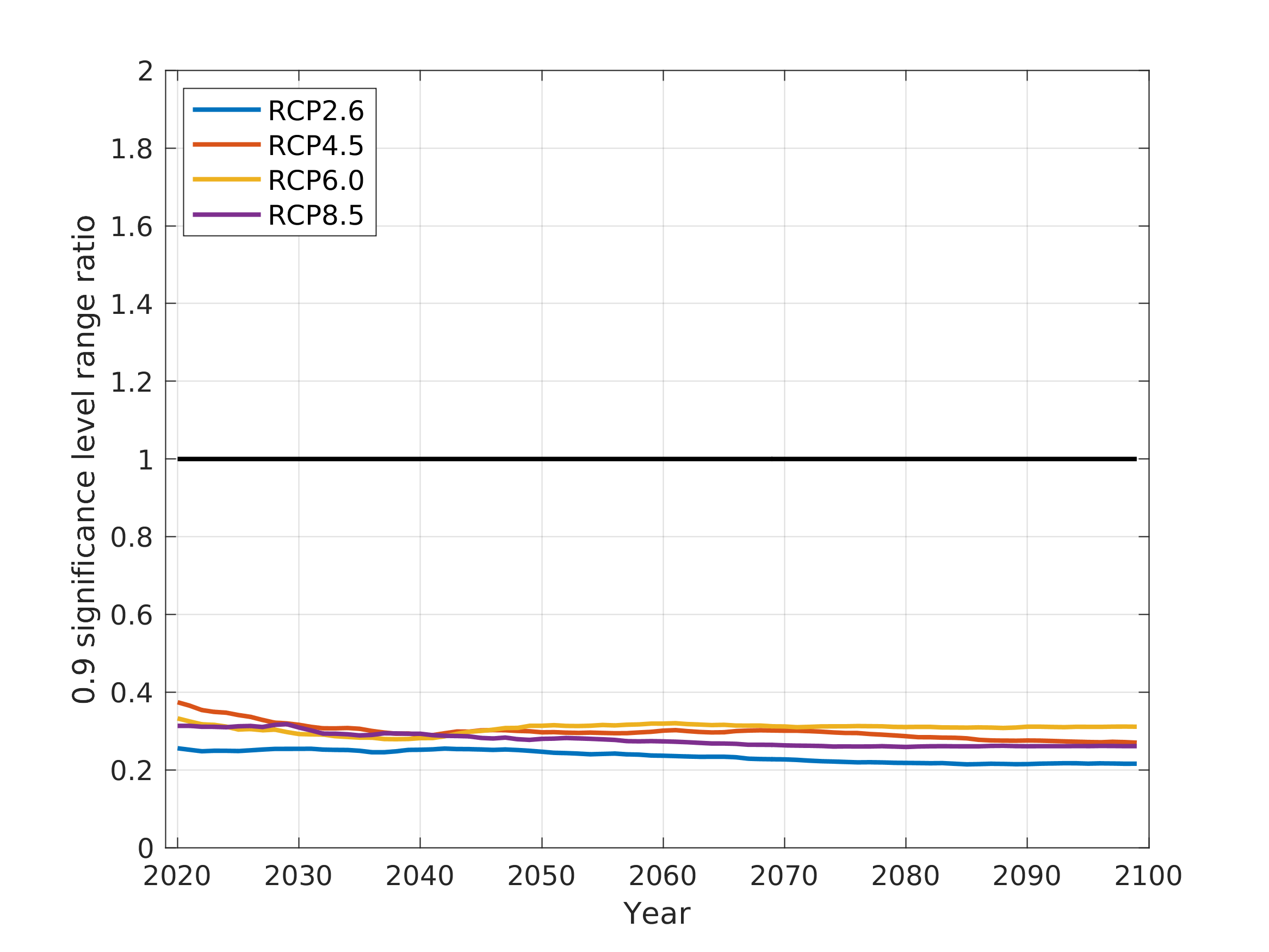}
\caption{\label{fig:cal_std_ratio_sym} The temporal variation of the ratio between the uncertainty range based on the EGA weighted ensemble and the AR method and the range based on the equally weighted ensemble and the Gaussian assumption for the 0.9 significance level. The different lines correspond to the different RCPs.}
\end{center}
\end{figure}
 
 \setcounter{table}{0}
\renewcommand{\tablename}{Extended Data Table}

\begin{table}
\centering
\caption{\label{quantile_AR/GS} The ratio between the projected GMST uncertainty ranges based on the EGA and AR and the equally weighted and Gaussian estimations. The values represent the ratio between the ranges for two different periods of 20 years as denoted. The different columns correspond to different RCPs, and the different rows correspond to different confidence levels.}
\begin{tabular} {| c | c | c | c | c | c | c | c | c |}
\hline
 & \multicolumn{4}{|c|}{$2046$--$2065$} & \multicolumn{4}{|c|}{$2080$--$2099$}\\
\hline
c & RCP2.6 & RCM4.5 & RCP6.0 & RCP8.5 & RCP2.6 & RCM4.5 & RCP6.0 & RCP8.5\\
\hline
0.9 & 0.23 & 0.3 & 0.32 & 0.27 & 0.22 & 0.27 & 0.31 & 0.26\\
0.8 & 0.18 & 0.25 & 0.25 & 0.22 & 0.17 & 0.23 & 0.24 & 0.22\\
0.7 & 0.19 & 0.27 & 0.26 & 0.26 & 0.17 & 0.24 & 0.26 & 0.26\\
0.6 & 0.16 & 0.23 & 0.21 & 0.2 & 0.15 & 0.21 & 0.21 & 0.2\\
0.5 & 0.17 & 0.21 & 0.24 & 0.17 & 0.16 & 0.19 & 0.24 & 0.17\\
0.4 & 0.18 & 0.24 & 0.27 & 0.21 & 0.17 & 0.22 & 0.27 & 0.2\\
0.3 & 0.18 & 0.27 & 0.3 & 0.25 & 0.16 & 0.24 & 0.29 & 0.24\\
0.2 & 0.18 & 0.3 & 0.22 & 0.23 & 0.16 & 0.27 & 0.21 & 0.23\\
0.1 & 0.24 & 0.35 & 0.2 & 0.33 & 0.22 & 0.32 & 0.19 & 0.33\\
\hline
 \end{tabular}
\end{table}

\begin{table}
\centering
\caption{\label{quantile_ranges} The uncertainty range of the projected GMST based on the EGA weighting and the AR estimation. The values correspond to two different periods of 20 years. The different columns correspond to different RCPs, and the different rows correspond to different confidence levels.}
\begin{tabular} {| c | c | c | c | c | c | c | c | c |}
\hline
 & \multicolumn{4}{|c|}{$2046$--$2065$} & \multicolumn{4}{|c|}{$2080$--$2099$}\\
\hline
c & RCP2.6 & RCM4.5 & RCP6.0 & RCP8.5 & RCP2.6 & RCM4.5 & RCP6.0 & RCP8.5\\
\hline
0.9 & 0.27 & 0.32 & 0.38 & 0.32 & 0.29 & 0.4 & 0.54 & 0.56\\ 0.8 & 0.16 & 0.21 & 0.23 & 0.21 & 0.18 & 0.26 & 0.33 & 0.36\\ 0.7 & 0.14 & 0.18 & 0.2 & 0.2 & 0.15 & 0.23 & 0.28 & 0.35\\ 0.6 & 0.1 & 0.13 & 0.13 & 0.12 & 0.1 & 0.16 & 0.19 & 0.22\\ 0.5 & 0.08 & 0.09 & 0.12 & 0.08 & 0.09 & 0.11 & 0.17 & 0.15\\ 0.4 & 0.07 & 0.08 & 0.1 & 0.08 & 0.07 & 0.1 & 0.15 & 0.14\\ 0.3 & 0.05 & 0.07 & 0.08 & 0.07 & 0.05 & 0.09 & 0.12 & 0.12\\ 0.2 & 0.03 & 0.05 & 0.04 & 0.04 & 0.03 & 0.06 & 0.06 & 0.07\\ 0.1 & 0.02 & 0.03 & 0.02 & 0.03 & 0.02 & 0.04 & 0.03 & 0.05\\
\hline
 \end{tabular}
\end{table}

\begin{table}
\centering
\caption{\label{quantile_ranges2} The uncertainty range of the projected GMST based on the equally weighted ensemble and the Gaussian estimation. The values correspond to two different periods of 20 years. The different columns correspond to different RCPs, and the different rows correspond to different confidence levels.}
\begin{tabular} {| c | c | c | c | c | c | c | c | c |}
\hline
 & \multicolumn{4}{|c|}{$2046$--$2065$} & \multicolumn{4}{|c|}{$2080$--$2099$}\\
\hline
c & RCP2.6 & RCM4.5 & RCP6.0 & RCP8.5 & RCP2.6 & RCM4.5 & RCP6.0 & RCP8.5\\
\hline
0.9 & 1.15 & 1.07 & 1.2 & 1.2 & 1.36 & 1.5 & 1.75 & 2.15\\
0.8 & 0.9 & 0.83 & 0.93 & 0.94 & 1.06 & 1.17 & 1.36 & 1.68\\
0.7 & 0.73 & 0.67 & 0.75 & 0.76 & 0.85 & 0.94 & 1.1 & 1.36\\
0.6 & 0.59 & 0.55 & 0.61 & 0.62 & 0.69 & 0.77 & 0.89 & 1.1\\
0.5 & 0.47 & 0.44 & 0.49 & 0.49 & 0.56 & 0.61 & 0.72 & 0.88\\
0.4 & 0.37 & 0.34 & 0.38 & 0.38 & 0.43 & 0.48 & 0.56 & 0.69\\
0.3 & 0.27 & 0.25 & 0.28 & 0.28 & 0.32 & 0.35 & 0.41 & 0.5\\
0.2 & 0.18 & 0.16 & 0.18 & 0.19 & 0.21 & 0.23 & 0.27 & 0.33\\
0.1 & 0.09 & 0.08 & 0.09 & 0.09 & 0.1 & 0.11 & 0.13 & 0.16\\
\hline
 \end{tabular}
\end{table}

\begin{table}
\centering
\caption{\label{gammas} $\gamma_u$ and $\gamma_d$ for different RCPs and significance levels.}
\begin{tabular} {| c | c | c | c | c | c | c | c | c |}
\hline
 & \multicolumn{2}{|c|}{RCP2.6} & \multicolumn{2}{|c|}{RCP4.5}& \multicolumn{2}{|c|}{RCP6.0} & \multicolumn{2}{|c|}{RCP8.5}\\
\hline
c & $\gamma_u$ & $\gamma_d$ & $\gamma_u$ & $\gamma_d$ & $\gamma_u$ & $\gamma_d$ & $\gamma_u$ & $\gamma_d$\\
\hline
0.9 & 0.591 & 0.227 & 0.701 & 0.307 & 0.76 & 0.29 & 0.653 & 0.293\\
0.8 & 0.318 & 0.181 & 0.407 & 0.25 & 0.45 & 0.19 & 0.395 & 0.215\\
0.7 & 0.264 & 0.152 & 0.343 & 0.232 & 0.37 & 0.18 & 0.38 & 0.202\\
0.6 & 0.157 & 0.136 & 0.203 & 0.195 & 0.21 & 0.15 & 0.192 & 0.174\\
0.5 & 0.136 & 0.109 & 0.168 & 0.118 & 0.19 & 0.14 & 0.143 & 0.103\\
0.4 & 0.107 & 0.094 & 0.15 & 0.108 & 0.16 & 0.13 & 0.131 & 0.1\\
0.3 & 0.084 & 0.062 & 0.12 & 0.093 & 0.14 & 0.09 & 0.122 & 0.082\\
0.2 & 0.064 & 0.032 & 0.107 & 0.048 & 0.06 & 0.05 & 0.072 & 0.054\\
0.1 & 0.042 & 0.023 & 0.09 & 0.001 & 0.03 & 0.02 & 0.051 & 0.039\\
\hline
 \end{tabular}
\end{table}

\begin{table}
\centering
\caption{\label{skewkurt} Skewness and kurtosis of the estimated (based on the EGA weighted ensemble and the AR estimation of the uncertainty) probability distribution of the 20-year average projected GMST. The cumulants are the same for all the years in the prediction period because the different years only differ in their average (which does not affect the central moments) and STD (which only sets the scale and does not affect the ratio between the moments of the probability distribution). The values of the correction factors $\gamma_{u,d}(c)$ are constant during the prediction period.}
\begin{tabular} {| c | c | c | c | c |}
\hline
& skewness & excess kurtosis\\
\hline
{RCP26} & 1.22 & 1.48\\
{RCP45} & 0.95 & 0.73\\
{RCP60} & 1.41 & 1.78\\
{RCP85} & 1.03 & 0.8\\
\hline
 \end{tabular}
\end{table}

\renewcommand{\arraystretch}{0.6}

\begin{table}
\footnotesize
\caption{\label{Models_table} Model Availability. Not all the models included in the CMIP5 projection data spanned the period of $1948$--$2100$ considered in our study, and most of the models did not provide projections for all the RCPs. The table lists the models included in our ensemble for each RCP.}
\begin{center}
\begin{tabular}{ p{4cm}  c c c c }
  \hline
 Model Name & RCP2.6 & RCP4.5 & RCP6.0 & RCP8.5 \\
  \hline
 ACCESS1-0      & X       & $\surd$ & $\surd$ & $\surd$ \\
 ACCESS1-3      & X       & $\surd$ & X       & $\surd$ \\
 CCSM4          & $\surd$ & $\surd$ & $\surd$ & $\surd$ \\
 CMCC-CESM      & X       & X       & X       & $\surd$ \\
 CMCC-CMS       & X       & $\surd$ & X       & $\surd$ \\
 CMCC-CM        & X       & $\surd$ & X       & $\surd$ \\
 CNRM-CM5       & $\surd$ & $\surd$ & X       & $\surd$ \\
 CSIRO-Mk3-6-0  & $\surd$ & $\surd$ & $\surd$ & $\surd$ \\
 CanESM2        & $\surd$ & $\surd$ & X       & $\surd$ \\
 GISS-E2-H-CC   & X       & $\surd$ & X       & $\surd$ \\
 GISS-E2-H      & $\surd$ & $\surd$ & $\surd$ & $\surd$ \\
 GISS-E2-R-CC   & X       & $\surd$ & X       & $\surd$ \\
%GISS-E2-R      & X       & X       & X       & X       \\
 HadGEM2-AO     & $\surd$ & $\surd$ & X       & $\surd$ \\
 HadGEM2-CC     & X       & $\surd$ & X       & $\surd$ \\
 HadGEM2-ES     & $\surd$ & $\surd$ & X       & $\surd$ \\
 IPSL-CM5A-LR   & $\surd$ & $\surd$ & $\surd$ & $\surd$ \\
 IPSL-CM5B-LR   & X       & $\surd$ & X       & $\surd$ \\
 MIROC-ESM-CHEM & $\surd$ & $\surd$ & $\surd$ & $\surd$ \\
 MIROC-ESM      & $\surd$ & $\surd$ & $\surd$ & $\surd$ \\
 MIROC5         & $\surd$ & $\surd$ & $\surd$ & $\surd$ \\
 MPI-ESM-LR     & $\surd$ & $\surd$ & X       & $\surd$ \\
 MPI-ESM-MR     & $\surd$ & $\surd$ & X       & $\surd$ \\
 MRI-CGCM3      & $\surd$ & $\surd$ & $\surd$ & $\surd$ \\
 MRI-ESM1       & X       & X       & X       & $\surd$ \\
 NorESM1-ME     & $\surd$ & $\surd$ & $\surd$ & $\surd$ \\
 NorESM1-M      & $\surd$ & $\surd$ & $\surd$ & $\surd$ \\
 CESM1-BGC      & X       & $\surd$ & X       & $\surd$ \\
 CESM1-CAM5     & $\surd$ & $\surd$ & $\surd$ & $\surd$ \\
 EC-EARTH       & X       & $\surd$ & X       & $\surd$ \\
 FGOALS-g2      & $\surd$ & $\surd$ & X       & $\surd$ \\
 GFDL-CM3       & $\surd$ & $\surd$ & $\surd$ & $\surd$ \\
 GFDL-ESM2G     & $\surd$ & $\surd$ & $\surd$ & $\surd$ \\
 GFDL-ESM2M     & $\surd$ & $\surd$ & $\surd$ & $\surd$ \\
  \hline
 Total          & 21      & 31      & 15      & 33      \\
 \end{tabular}
\begin{flushleft}
\end{flushleft}
\end{center}
\end{table}

\begin{table}
\footnotesize
\caption{\label{Models_weights} The EGA weight assigned to each model and for each ensemble (different ensembles for different RCPs). These weights, assigned at the end of the learning period, remain time-independent during the projection period.}
\begin{center}
\begin{tabular}{ p{4cm}  c c c c }
  \hline
 Model Name & RCP2.6 & RCP4.5 & RCP6.0 & RCP8.5 \\
  \hline
{ACCESS}{1} & X & 0.02 & 0.02 & 0.02\\
{ACCESS}{1}-3 & X & 0.04 & X & 0.04\\
{CCSM}{4} & 0.02 & 0.01 & 0.07 & 0.01\\
{CMCC}-{CESM} & X & X & X & 0.03\\
{CMCC}-{CMS} & X & 0.01 & X & 0.01\\
{CMCC}-{CM} & X & 0.02 & X & 0.02\\
{CNRM}-{CM}{5} & 0.03 & 0.02 & X & 0.02\\
{CSIRO}-{Mk}{3}-6 & 0.04 & 0.05 & 0.03 & 0.03\\
{CanESM}{2} & 0.01 & X & X & X\\
{GISS}-E{2}-{CC}-H & X & 0.02 & X & 0.02\\
{GISS}-E{2}-H & 0.06 & 0.04 & 0.08 & 0.04\\
{GISS}-E{2}-{CC}-R & X & 0.03 & X & 0.03\\
{GISS}-E{2}-R & X & X & X & X\\
{HadGEM}{2}-{AO} & 0.01 & 0.01 & X & 0.01\\
{HadGEM}{2}-{CC} & X & 0.02 & X & 0.02\\
{HadGEM}{2}-{ES} & 0.02 & 0.01 & X & 0.01\\
{IPSL}-{CM5A}-{LR} & 0.01 & X & 0.02 & X\\
{IPSL}-{CM5B}-{LR} & X & 0.06 & X & 0.05\\
{MIROC}-{ESM}-{CHEM} & 0.07 & 0.05 & 0.11 & 0.05\\
{MIROC}-{ESM} & 0.03 & 0.02 & 0.05 & 0.02\\
{MIROC}{5} & 0.02 & 0.01 & 0.01 & 0.01\\
{MPI}-{LR}-{ESM} & 0.06 & 0.03 & X & 0.03\\
{MPI}-{ESM}-{MR} & 0.02 & 0.01 & X & 0.01\\
{MRI}-{CGCM}{3} & 0.25 & 0.25 & 0.24 & 0.21\\
{MRI}-{ESM}{1} & X & X & X & 0.08\\
{NorESM}{1}-{ME} & 0.06 & 0.04 & 0.08 & 0.04\\
{NorESM}{1}-M & 0.04 & 0.03 & 0.08 & 0.03\\
{CESM}{1}-{BGC} & X & 0.01 & X & 0.01\\
{CESM}{1}-{CAM}{5} & 0.09 & 0.07 & 0.12 & 0.06\\
{EC}-{EARTH} & X & 0.02 & X & 0.01\\
{FGOALS}-g{2} & 0.08 & 0.04 & X & 0.04\\
{GFDL}-{CM}{3} & 0.01 & 0.01 & 0.01 & 0.01\\
{GFDL}-{ESM2G} & 0.01 & 0.01 & 0.03 & 0.01\\
{GFDL}-{ESM2M} & 0.03 & 0.02 & 0.05 & 0.02 \\
\hline
 Total          & 1      & 1      & 1      & 1      \\
 \end{tabular}
\begin{flushleft}
\end{flushleft}
\end{center}
\end{table}

\renewcommand{\arraystretch}{1}

\begin{table}
\centering
\caption{\label{quantiles_20yr} Quantiles for the 20-year average projected GMST. Values in parentheses were taken from the last IPCC report\cite{IPCC_2013}. The quantiles are based on the EGA weighted ensemble and the AR estimation of the uncertainties. For all the RCPs and for both periods, we estimate a narrower distribution than the corresponding IPCC estimation.}
\begin{tabular} {| c | c | c | c | c | c | c  |}
  \hline
& \multicolumn{3}{|c|}{$2046$--$2065$} & \multicolumn{3}{|c|}{$2080$--$2099$ ($2081$--$2100$)} \\
\hline
& $5\%$ & $50\%$ & $95\%$ & $5\%$ & $50\%$ & $95\%$ \\
\hline
{RCP2.6} & 0.92 (0.4) & 0.99 (1.0) & 1.19 (1.6) & 0.96 (0.3) & 1.04 (1.0) & 1.26 (1.7) \\
{RCP4.5} & 1.28 (0.9) & 1.37 (1.4) & 1.6 (2.0) & 1.71 (1.1) & 1.82 (1.8) & 2.11 (2.6) \\
{RCP6.0} & 1.16 (0.8) & 1.26 (1.3) & 1.54 (1.8) & 2.08 (1.4) & 2.23 (2.2) & 2.63 (3.1) \\
{RCP8.5} & 1.86 (1.4) & 1.95 (2.0) & 2.18 (2.6) & 3.37 (2.6) & 3.53 (3.7) & 3.92 (4.8) \\
\hline
 \end{tabular}
\end{table}

\begin{table}
\centering
\caption{\label{quantiles_ranges_20yr} Ranges from the $5\%$--$95\%$ quantiles for the 20-year average projected GMST. The range estimated using the EGA weighted ensemble and the AR method for estimating the uncertainties is considerably smaller than the IPCC reported range for all the RCPs. }
\begin{tabular} {| c | c | c |}
  \hline
& \multicolumn{2}{|c|}{$5\%$--$95\%$} \\
\hline
& \multicolumn{1}{|c|}{$2046$--$2065$} & \multicolumn{1}{|c|}{$2080$--$2099$ ($2081$--$2100$)} \\
\hline
{RCP2.6} & 0.27 (1.2) & 0.30 (1.4) \\
{RCP4.5} & 0.32 (1.1) & 0.40 (1.5) \\
{RCP6.0} & 0.38 (1.0) & 0.55 (1.7) \\
{RCP8.5} & 0.32 (1.2) & 0.55 (2.2) \\
\hline
 \end{tabular}
\end{table}

\end{addendum}

\end{document}